\documentclass[sigconf]{acmart}

\usepackage{listings}
\usepackage{algorithm}
\usepackage{algorithmic}
\usepackage{xspace}

\newcommand{\vecr}{\mathbf{r}}
\newcommand{\vecR}{\mathbf{R}}
\newcommand{\code}[1]{\texttt{#1}}
\newcommand{\trash}[1]{}

\newcommand{\eg}[0]{e.\,g.,\xspace}

\newcommand{\registered}{\textsuperscript{\textregistered}\xspace}
\newcommand{\trademark}{\texttrademark\xspace}

\newcommand{\xphiTMKNL}[0]{Intel\registered Xeon Phi\trademark processor\xspace}

\newcommand{\xeonReg}[0]{Intel\registered Xeon\registered}

\newcommand{\vtuneTM}{Intel\registered VTune\trademark Amplifier 2017\xspace }

\newcommand{\studioReg}{Intel\registered Parallel Studio XE 2017\xspace }

\newcommand{\CompilerDisclaimer}{ Optimization Notice: Intel's compilers may or
  may not optimize to the same degree for non-Intel microprocessors for
  optimizations that are not unique to Intel microprocessors. These
  optimizations include SSE2, SSE3, and SSSE3 instruction sets and other
  optimizations. Intel does not guarantee the availability, functionality, or
  effectiveness of any optimization on microprocessors not manufactured by
  Intel. Microprocessor-dependent optimizations in this product are intended
  for use with Intel microprocessors. Certain optimizations not specific to
  Intel microarchitecture are reserved for Intel microprocessors. Please refer
  to the applicable product User and Reference Guides for more information
regarding the specific instruction sets covered by this notice. }

\newcommand{\BenchmarkDisclaimer}{ Software and workloads used in performance
  tests may have been optimized for performance only on Intel microprocessors.
  Performance tests, such as SYSmark and MobileMark, are measured using
  specific computer systems, components, software, operations and functions.
  Any change to any of those factors may cause the results to vary. You should
  consult other information and performance tests to assist you in fully
  evaluating your contemplated purchases, including the performance of that
  product when combined with other products.   For more complete information
  visit www.intel.com/benchmarks.  
\\ Intel, Xeon, and Intel Xeon Phi are trademarks of Intel Corporation in the
U.S. and/or other countries.  }

\DeclareMathVersion{sans}
\lstnewenvironment{sflisting}[1][]
{\lstset{#1}\mathversion{sans}}{}

\begin{document}

\title{Embracing a new era of highly efficient and productive quantum Monte Carlo simulations}

\author{Amrita Mathuriya}
\affiliation{
  \institution{Intel Corporation}
}
\email{amrita.mathuriya@intel.com}

\author{Ye Luo}
\affiliation{
  \institution{Argonne National Laboratory}
}
\email{yeluo@anl.gov}

\author{Raymond C. Clay III }
\affiliation{
  \institution{Sandia National Laboratories}
}
\email{rclay@sandia.gov}

\author{Anouar Benali}
\affiliation{
  \institution{Argonne National Laboratory}
}
\email{benali@anl.gov}

\author{Luke Shulenburger}
\affiliation{
  \institution{Sandia National Laboratories}
}
\email{lshulen@sandia.gov}

\author{Jeongnim Kim}
\affiliation{
  \institution{Intel Corporation}
}
\email{jeongnim.kim@intel.com}

\begin{abstract}
  QMCPACK has enabled cutting-edge materials research on supercomputers for over
a decade. It scales nearly ideally but has low single-node efficiency due to
the physics-based abstractions using array-of-structures objects, causing
inefficient vectorization. We present a systematic approach to transform
QMCPACK to better exploit the new hardware features of modern CPUs in portable
and maintainable ways. We develop miniapps for fast prototyping and
optimizations.  We implement new containers in structure-of-arrays data layout
to facilitate vectorizations by the compilers.  Further speedup and smaller
memory-footprints are obtained by computing data on the fly with the vectorized
routines and expanding single-precision use.  All these are seamlessly
incorporated in production QMCPACK.  We demonstrate upto 4.5x speedups on
recent Intel\registered processors and IBM Blue Gene/Q for representative workloads.
Energy consumption is reduced significantly commensurate to the speedup factor.
Memory-footprints are reduced by up-to 3.8x, opening the possibility to solve
much larger problems of future. 

\end{abstract}

\maketitle


\section{Introduction}

Large-scale parallel computing resources have enabled numerous science
discoveries and grand-challenge simulations since the early 1990s.  Productive
utilization of high-performance computing (HPC) resources demands
algorithms and implementations that are both highly efficient and scalable.
The gap between the peak and sustained performance that a typical HPC application
can achieve has been steadily growing. The news article ``4 applications
sustain 1 petaflop on Blue Waters'' in 2013~\cite{bwpeta} manifests the
challenges the developers are facing to exploit the powerful systems at scale.  

\begin{figure}
  \centering
  \includegraphics[scale=0.38]{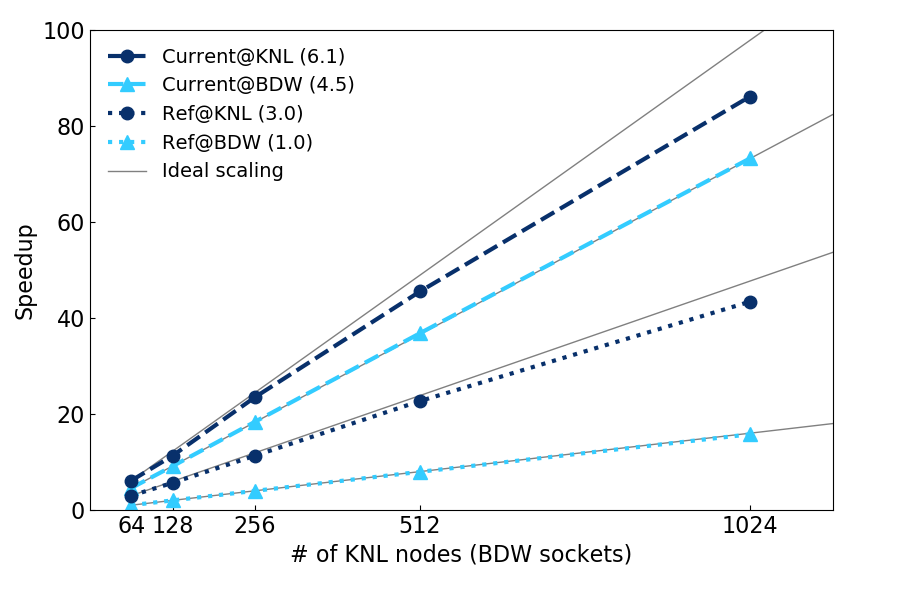}
  \caption{Strong scaling of NiO-64 benchmark on Trinity at LANL (KNL) and
  Serrano at SNL (BDW) systems.  The performance is normalized by a reference
throughput using 64 BDW sockets.  Slopes of the ideal-scaling lines are provided
in parentheses.}
  \label{fig:scaling}
\end{figure}

Multiple factors are responsible for the growing performance gap.  The
increasing complexity of HPC applications, a fast evolving hardware
landscape, and a wide range of programming models offered to the developers ---
all play roles in the decreasing productivity of extremely capable HPC systems.
Lately, much of the increase in computing power of the processors comes from
increasing opportunities for parallelism on a node through many cores, multiple
hardware threads and wide SIMD units. Without fully exploiting these
parallelisms and unique hardware features, such as high-bandwidth memory and
cache subsystems, applications leave a lot of potential performance gain on the
table.  However, any change of a production-level HPC application to adopt and
adapt to new HPC infrastructure is a formidable task even for a team of highly
experienced developers.  This work explores portable and maintainable methods to
transform QMCPACK~\cite{qmcpackSite}, which has similar compute and design
characteristics to many HPC applications, to achieve significantly more
efficient single-node performance.

Quantum Monte Carlo (QMC) is a highly accurate, but computationally demanding method.
It consumes a significant
fraction of US-DOE resources every year, leveraging highly scalable algorithms and
implementations. Typical QMC calculations use 1000s of nodes at a time on the
leadership facilities.  QMCPACK implements hybrid parallelism with OpenMP and
MPI~\cite{qmcpack} and has close to ideal parallel efficiency as shown in
Fig.~\ref{fig:scaling}.  The figure shows strong scaling of NiO-64 benchmark on
2nd generation Intel\registered Xeon Phi\trademark processor (KNL) and
Intel\registered Xeon\registered E5v4 processor (BDW).  However, on-node
efficiency is low and it achieves below 10\% of the peak performance even on
Blue Waters~\cite{bwpeta}.
Compounding this problem, it does not utilize SIMD parallelism to the fullest
extent except for special kernels using platform-dependent intrinsics, \eg
QPX intrinsics \cite{YeSC15} on IBM BG/Q, or SSE/SSE2 intrinsics on x86.  As shown
in Fig.~\ref{fig:scaling}, our work on QMCPACK increases on-node efficiency by
2-4.5x, which translates directly to a multi-node speedup of the same factor,
with nearly ideal scaling. As an added benefit, this increase in compute efficiency
impacts not only scientific productivity, but also 
results in similar improvement in energy efficiency. 

This work presents a systematic approach to transform QMCPACK.  We use 
representative workloads of various problem sizes and computational
characteristics on multiple platforms to develop a set of miniapps to optimize
the most computationally expensive components of the application.  The use of
miniapps facilitates exploration of a large design space and algorithms and
fast prototyping of new methods while maintaining realistic code usage. The
full integration of the new solutions is then staged to evaluate the
performance impact of each step, minimize the changes in the high-level QMC
drivers and validate the correctness of the implementations. We analyze the
performance evolution throughout the optimization processes and iteratively
improve both miniapps and the full application. 

Based on the extensive performance analysis of the current workloads including
those used in this work, we set two main targets to increase the performance:
i) improve SIMD efficiency and ii) reduce memory footprint.  We aim to develop
portable and maintainable solutions to increase the productivity of the QMC
experts who use QMCPACK to develop new electronic structure theories, numerical
techniques and parallel algorithms. Hence, the code transformations are
constrained to use C++11 and OpenMP 4 standards, consistent with the existing
physics abstractions and the thread-level parallelization in QMCPACK. No
platform-specific optimizations are employed for this work. However, the
infrastructure --- miniapps, classes/interfaces \textit{etc} --- is devised to be
extensible.  Specialization for a specific hardware can be added for further
improvement.

We demonstrate the performance impact using four representative workloads.  Our
work speeds up QMCPACK simulations by 2-4.5x on KNL and BDW clusters of up to
1024 MPI tasks as Fig.~\ref{fig:scaling} shows.  The energy usage reduction in
proportion to the speedup factor on KNL system is achieved by the optimizations
of this work. The memory usage is reduced to fit in KNL's 16GB MCDRAM memory
for a large problem with 784 electrons.  Our work leads to more productive QMC
simulations by enabling users to solve larger problems quicker.

%

\begin{figure}
  \centering
  \includegraphics[scale=0.35]{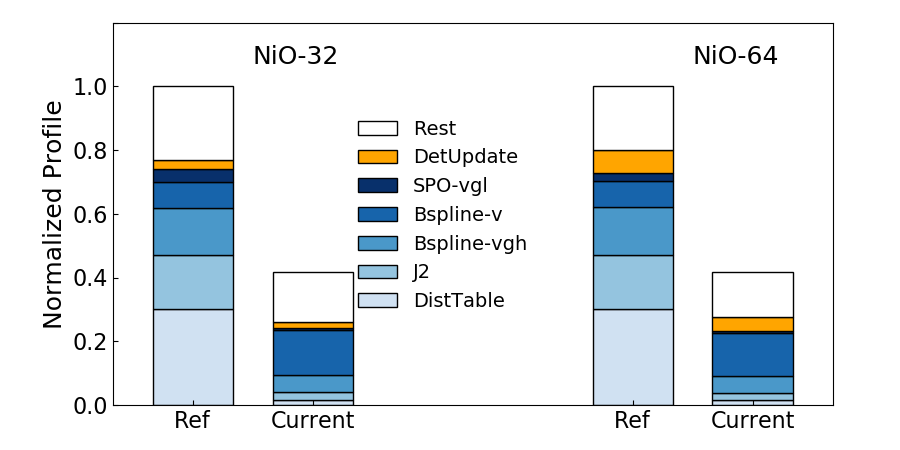}
  \caption{Normalized hot-spot profiles on KNL. 
  Current version profiles accomodate the speedup wrt. Ref version for the corresponding benchmark. }
  \label{fig:prof}
\end{figure}

\subsection{Summary of work and contributions}

A detailed analysis of the latest release reveals low SIMD efficiency in key
compute kernels. The top hot-spots of the reference profiles in
Fig.~\ref{fig:prof} are DistTable, J2 and Bspline which use
array-of-structures (AoS) data types to represent 3D physics of the electrons,
such as the positions of $N$ electrons in \code{R[N][3]}. This abstraction is
the foundation for all high-level algorithms in QMCPACK.  However, highly
productive abstractions for science can incur a high abstraction penalty as
demonstrated by numerous studies~\cite{Veldhuizen2004,Muller2000}.  For this
reason, we introduce complementary objects of structure-of-arrays (SoA) types
for all the compute expensive kernels. For example,
\code{Rsoa[3][N]} for \code{R} is added to enable efficient vectorization and
increase bandwidth utilization. 

Also, the memory footprint of QMCPACK grows as $\mathcal{O}(N^2)$ with the
number of electrons and can quickly become challenging.  Fully utilizing a KNL
system requires large number of threads and walkers (samples), making the footprints
larger, compared to regular Intel\registered Xeon\registered systems.  We solve
the problem with i) mixed-precision (MP) and ii) compute-on-the-fly algorithms.  By expanding
single precision use in the key data structures and methods, we reduce memory use
and bandwidth demands as well as making the computations faster.  Once the key
computational kernels become faster by exploiting efficient vectorization, it
becomes faster to compute elements when they are used than to store and
retrieve them.  These changes result in much decreased run time and better
memory usage. 

Transforming a big application of millions of lines of code and 1000s of files
to adopt new data layouts and mixed-precision algorithms is a big and
complicated task and must be carried out carefully to improve both application
performance and science productivity. The unique properties of QMC
algorithms and the object-oriented and generic framework of QMCPACK are
exploited to transform the full application.  New SoA objects are added to
improve the SIMD efficiency in the critical routines and the existing
abstractions and AoS objects are reused.  The miniapps facilitate fast
prototyping and evaluations and minimize the risk of the global transformations
until they are proven to be effective in realistic QMC simulations.

\noindent
In summary, \textbf{Contributions} of this publication are following: 
\begin{itemize}
  \item Created miniapps representing compute and 
    data access patterns of QMC simulations and used them to integrate
    the new developments to the full production QMCPACK.
  \item Implemented SoA data types, facilitating efficient vectorization of all
    the compute intensive kernels using C++11 and OpenMP standards. 
  \item Developed forward update and compute-on-the-fly algorithms, enabling
    further speedup and memory footprint reduction.
  \item Expanded single-precision use in CPU code for memory reduction and speed
and improved the accuracy of the mixed-precision methods for both CPU and GPU ports.
\end{itemize}

\section{Related work}

Microkernels or miniapps have been widely used for HPC procurements or
acceptance testing. For instance, the CORAL microkernel benchmarks are code
snippets extracted from HPC applications.  These are intended to address
certain capabilities of a system, such as NEKbonemk used for SIMD compiler
challenge~\cite{coral}. The miniapps of this work are intended to
spur QMC development, going beyond the traditional roles of microkernels. 
They reproduce the computational patterns, memory usage, data access and
thread-level parallelism of the full code as realistically as possible.
Performance changes in these miniapps are reliable predictors of the
performance of real QMC simulations.  We use them to narrow the solution space
for the optimization and parallelization of QMCPACK.

Our previous work~\cite{qmcipdps17} showed performance improvement in 3D B-spline routines
using a SoA data type.  In this work, we implement the SoA data types in the
full QMCPACK code for the top kernels. Also, we use generic C++ containers to 
port the optimizations instead of using plain old data types.

\code{VectorSoaContainer<T,D>} (VSC) adopts the concepts of SIMD Data Layout
Templates (SDLT) library introduced in Intel\registered C++ Compiler
17.0~\cite{sdlt,sdltqlib}. VSC is a generic SoA container of \code{C[D][N]} for
D dimensional particle simulations, providing access operators and methods. 
Current SDLT only supports ``plain old data'' objects and adopting it in
QMCPACK would require large-scale refactoring while losing the generality it
aims to maintain.  Therefore, we introduce VSC to express the high-level QMC
algorithms as before, while hiding the implementation details --- memory allocation and
layout.  Very limited changes are made at the physics abstraction level and
they are mostly to incorporate new algorithms.

Single precision has been extensively used in QMCPACK's GPU
port~\cite{qmcpackGPU} resulting in significant speedups and memory savings.
Single precision was later introduced to the CPU version to compute the 3D
B-spline SPOs (single-particle orbitals)~\cite{qmcpackman}. 
This work expands the use of single precision
to the entire QMC calculations.  To preserve numerical accuracy for both CPU
and GPU ports, the quantities per walker and for the ensemble are computed in
double precision and new states are periodically computed from
scratch~\cite{MixedPrecision}. 

QMCPACK makes extensive use of object-oriented and generic programming and
design patterns~\cite{patterns} for reusability and
extensibility~\cite{qmcpack}.  Computational efficiency is achieved through
inlined specializations of C++ template and by using SIMD intrinsics for core
kernels~\cite{qmcpackman,qmcpackGPU}.  This work eliminates the
platform-dependent optimization and leverages optimizing C++ compilers and
OpenMP standards to achieve greater efficiency on modern CPUs. Many features in
C++11~\cite{cpp11} are used to make the code compact, efficient and
maintainable.

\section{QMC algorithms}

In quantum mechanics, all physically observable quantities for a system
containing $N$ particles can be computed from the $3N$-dimensional {\em wave
function}, $\Psi(\mathbf{r}_1,\dots,\mathbf{r}_{N})$~\cite{foulkes}.  For any
trial {\em wave function}, $\Psi_T(\mathbf{R})$, we can compute an energy as
the expectation value of the many-body Hamiltonian, $\hat{H}$, 
\begin{equation} E_T = 
\frac{\int d^{3N}\mathbf{R} \ \Psi_T^*(\mathbf{R}) \hat{H}
\Psi_T(\mathbf{R})} {\int d^{3N}\mathbf{R} \ \left|\Psi_T(\mathbf{R})\right|^2}
\label{eq:Etrial}, 
\end{equation} 
where $\mathbf{R}$ is a $3N$-dimensional vector representing the positions of
the $N$ particles.  The direct evaluation of many-dimensional integrals of
Eq.~(\ref{eq:Etrial}) by stochastic sampling enables us to employ highly
accurate variational wave functions which can capture crucial many-body effects
in an efficient manner.  The Slater-Jastrow trial wave function used in this
work is
\begin{equation}
\Psi_T= \exp (J) D^{u} (\{\phi\}) D^{d} (\{\phi\}),
\label{eq:sj}
\end{equation}
with $N=N^{u}+N^{d}$ for the up and down spins.  For the rest of the paper, we
assume $N^u=N^d=N/2$.

\begin{figure} \centering
  \includegraphics[scale=0.38]{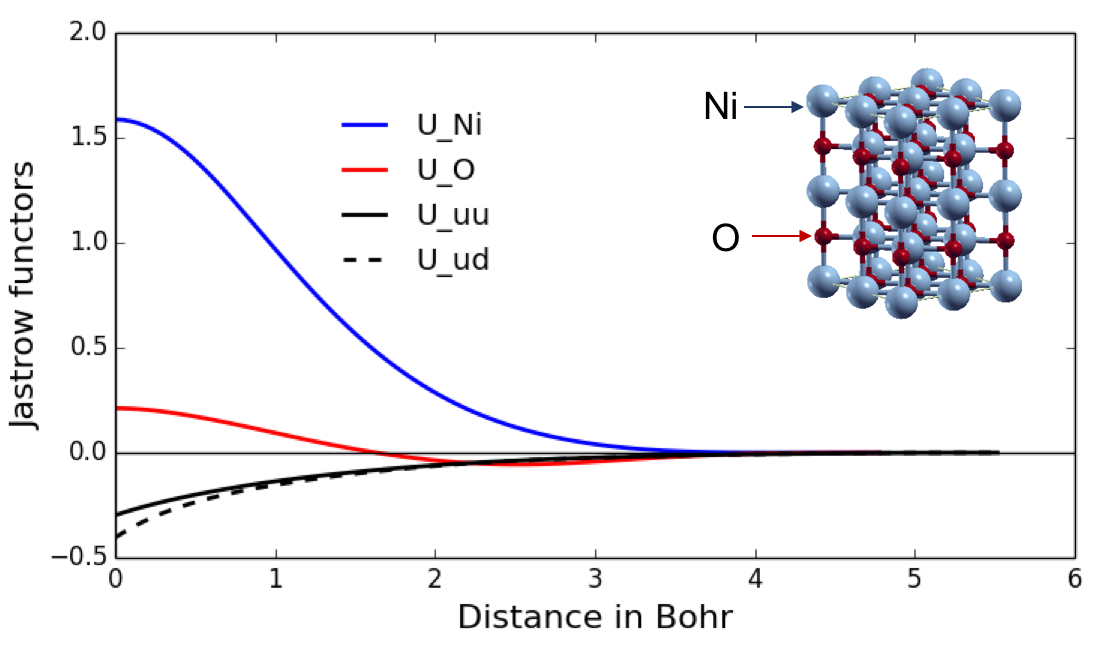} 
  \caption{Jastrow functors of Ni and O ions and up and down electron spins
   for a 32-atom supercell of NiO.}
  \label{fig:nioj} 
\end{figure}

The Jastrow factor $J$ describes the dynamic correlation and is factorized into
one-body, two-body and high-order correlation functions as
\begin{equation}
  J = \sum_I^{N_{\rm ion}}\sum_i^{N} U_I(|{\bf r}_I-{\bf r}_i|) 
  + \sum_{j \ne i}^{N} U_2(|{\bf r}_i-{\bf r}_j|) + \cdots
  \label{eq:j}
\end{equation}
Figure~\ref{fig:nioj} shows distinct Jastrow functors optimized for a
32-atom supercell of NiO.  The one-dimensional cubic B-spline is extensively used in QMCPACK
because of its generality and computational efficiency~\cite{bspline}.  

The Slater determinant captures the static correlation and ensures the
antisymmetric property of a Fermionic wave function upon exchange of a pair of
electrons as $D=\det|\mathbf{A}|$ and ${A}(i,j)=\phi_i(\vecr_j)$.  Here,
$\{\phi\}$ denotes a set of SPOs, often taken to be
the solution of a mean-field method such as density functional theory or the
Hartree-Fock approximation.

The diffusion Monte Carlo algorithm (DMC) shown in Alg.~\ref{alg:DMC} is the
most time-consuming stage of a QMC exploration of a system. We can define the
efficiency of a DMC calculation as $\kappa = 1/(\sigma^2 \tau_{\rm corr} T_{\rm MC})$,
where $\sigma$ is the variance for the optimized $\Psi_T$.  Increasing
computational and parallel efficiency impacts the DMC efficiency by reducing
the total MC time $T_{\rm MC}$ to reach the target statistical error.  The
auto-correlation time $\tau_{\rm corr}$~\cite{autocorr} reflects the quality of
$\Psi_T$ and the MC algorithms.  The ensemble size, the average number of
walkers, is important to reduce systematic errors due to the finite time step
and population.

\begin{algorithm}
  \begin{algorithmic}[1]
    \FOR{$\text{MC generation}=1\cdots M$}
      \FOR{$\text{walker}=1\cdots N_w$}
        \STATE let ${\bf R} = \{{\bf r}_1 \ldots {\bf r}_{N}\}$
        \FOR{$\text{particle}\  k=1 \cdots N$}
           \STATE set ${\bf r}_k^{\prime} \leftarrow {\bf r}_k+\nabla_k \Psi_T({\bf R)}+\delta$
           \STATE let ${\bf R}^{\prime} = \{{\bf r}_1 \ldots {\bf r}_k^{\prime} \ldots{\bf r}_{N}\}$
           \STATE \textbf{ratio} $\rho = \Psi_T ({\bf R}^{\prime})/\Psi_T ({\bf R})$
           \STATE \textbf{derivatives} $\nabla_k \Psi_T({\bf R}^{\prime}), \nabla^2_k\Psi_T({\bf R}^{\prime})$
           \STATE Accept $\vecr_k \leftarrow \vecr_k^{\prime}$ or reject
        \ENDFOR \COMMENT{particle}
        \STATE \textbf{local energy} $E_L=\hat{H}\Psi_T({\bf R})/\Psi_T({\bf R})$ 
      \ENDFOR \COMMENT{walker}
      \STATE reweight and branch walkers 
      \STATE update $E_T$ and load balance
    \ENDFOR \COMMENT{MC generation}
  \end{algorithmic}
\caption{Pseudocode for diffusion Monte Carlo.\label{alg:DMC}}
\end{algorithm}

A typical DMC implementation employs a particle-by-particle (PbyP) update for
the \textit{drift-and-diffusion stage} (L4-L10) to increase the MC efficiency.
Only one particle is moved at a time in this algorithm.  Once a new
configuration is sampled, the physical quantities, such as the local
energy $E_L$, are measured for the fixed electron positions.  The rest consists
of computing the trial energy $E_T$, taking statistics and load balancing of
the fluctuating population.

The Slater-Jastrow form used for the trial wavefunction, $\Psi_T$,
Eq.~(\ref{eq:sj}-\ref{eq:j}) is physically motivated but also has many
computational advantages for the PbyP update. Take the step of moving the
$k$-th electron from ${\bf r}_k$ to ${\bf r}_k^{\prime}$. The computation of
the ratio becomes
\begin{equation}
  \frac{\Psi_T({\bf r}_1\cdots {\bf r}_k^{\prime} \cdots {\bf r}_{N})}
  {\Psi_T({\bf r}_1\cdots {\bf r}_k \cdots {\bf r}_{N})}
  = \exp^{\Delta J_1} \exp^{\Delta J_2} 
  \frac{\det|\mathbf{A}^{\prime}|}{\det|\mathbf{A}|}, 
  \label{eq:ratio}
\end{equation}
where
\begin{eqnarray}
  \Delta J_1 &= \sum_I^{N_{\rm ion}} U_I(|\vecr_I-\vecr_k^{\prime}|)-\sum_I^{N_{\rm ion}} U_I(|\vecr_I-\vecr_k|),\nonumber \\
  \Delta J_2 &= \sum_{i\neq k}^{N} U_2(|\vecr_i-\vecr_k^{\prime}|)-\sum_{i\neq k}^{N} U_2(|\vecr_i-\vecr_k|).
  \label{eq:Jratios}
\end{eqnarray}
The determinant ratio is a dot product of the $k$-th row of $\mathbf{A}^{-1}$
and $v(\phi_1(\vecr_k^{\prime}),\cdots,\phi_{N/2}(\vecr_k^{\prime}))$ using
\begin{equation} \label{md_lemma}
  \det (\mathbf{A} + u{e}_k') = (1 + {e}_k'\mathbf{A}^{-1}u)\det(\mathbf{A}).
\end{equation}
The derivatives for the quantum forces on the electron are evaluated using the
same matrix determinant lemma~\cite{fahy_variational_1990,ClarkJCP2011}.  When
the proposed $\vecr_k^{\prime}$ is accepted, $\mathbf{A}^{-1}$ is updated using
Sherman-Morrison formula.  Other internal states, such as the distance tables
between the electrons and ions for Jastrow computations are updated to proceed
to the next particle move. 

Once a new configuration is obtained, $E_L$ is computed as
\begin{equation}
  E_L=-\frac{\nabla^2 \Psi_T(\vecR)}{2{\Psi_T(\vecR)}} 
  + \sum_{i<j}\frac{1}{|\vecr_i-\vecr_j|}
+ \sum_{I}\frac{\hat{V}_{\rm NL}\Psi_T(\vecR)}{\Psi_T(\vecR)}.
\end{equation}
The non-local pseudopotential operator $\hat{V}_{\rm NL}$ is handled by
approximating an angular integral by a quadrature on a spherical shell
surrounding each ion~\cite{fahy_variational_1990}. This requires ratio
evaluations of the electrons within a cutoff radius of an ion using
~Eq.~(\ref{eq:ratio}).

\section{Science goals}

Whereas diffusion Monte Carlo has in the past been applied to calculate the
properties of idealized highly crystalline materials with high accuracy, the
additional computing power that will be brought to bear as supercomputing
pushes past the petascale and into the exascale will bring with it the
possibility of treating the complexity of realistic materials.  If properly
harnessed, this could enable new kinds of scientific problems to be addressed.
For example, it could be possible to study the aging of photovoltaic materials
exposed to the environment rather than just their performance in a laboratory.

In order to meet these goals, however, a code will have to cope with several
changing features of the exascale landscape.  Firstly, increasing
parallelization is arriving often in the form of ever wider vector units
instead of increasing numbers of computing cores and secondly, the memory per
core is not necessarily increasing at a pace that will satisfy quantum Monte
Carlo's $O(N^2)$ memory footprint.  To successfully deal with these hurdles, a
code will have to increase vectorization while being as conservative as
possible with memory utilization.

\subsection{Benchmark problems}

\begin{table}[!t]
\caption{Workloads used in this work and their key properties.}
\label{tab:wl}
\centering
\small
\begin{tabular}{|l|c|c|c|c|}
\hline
& Graphite & Be-64 & NiO-32 & NiO-64\\
\hline
$N$ & 256 & 256 & 384 & 768\\
\hline
$N_{\rm ion}$ & 64 & 64 & 32 & 64\\
\hline
$N_{\rm ion}$/uint cell & 4 & 2 & 4 & 4\\
\hline
\# of uint cells & 16 & 32 & 8 & 16\\
\hline
Ion types ($Z^{*}$)  & C (4)& Be (4) & \multicolumn{2}{c|}{Ni(18), O(6)} \\
\hline
\hline
\# of unique SPOs & 80 & 81 & 144 & 240 \\
\hline
FFT grid & 28x28x80 & 84x84x144 & \multicolumn{2}{c|}{80x80x80} \\
\hline
B-spline (GB) & 0.1 & 1.4 & 1.3 & 2.1\\
\hline
\end{tabular}
\end{table}

In this work we will consider four different benchmark systems in order to
demonstrate how these algorithmic improvements in QMCPACK have addressed these
challenges.  The first is a classic throughput based benchmark which was
included in the assessment criteria for the CORAL machines\cite{coral}.  That
benchmark requires calculating the energy of a crystalline domain of graphite,
the precursor material for generating graphene.  The second benchmark requires
the calculation of the properties of beryllium.  This system was chosen because
it has a similar number of electrons (and hence computational scaling) as the
graphite benchmark, but as it is a lighter element, it can be performed without
the use of pseudopotentials.  The pseudopotentials are a crucial algorithmic
consideration necessary for treating heavier elements, but from a computational
point of view, their use stresses parts of the algorithm that are not expected
to be as important as the size of the problem increases.

The final two benchmarks are closely related.  They perform calculations on
crystals of NiO, an electronically strongly correlated material that is
difficult to treat for many methods.  These benchmarks involve calculations on
32 and 64 atom supercells of NiO and provide the most direct assessment of the
sort of calculations that are expected to be scientifically important in the
near future.   Table \ref{tab:wl} summarizes key features of these four
benchmarks, including the numbers of electrons in each and the number of single
particle orbitals required to calculate the trial wavefunction for
each one.  As the number of electrons is the single most important
factor affecting the performance profile, in most of the discussion that follows
we will focus on only the NiO 32 and 64 atom benchmarks.  These cases
involve pseudopotentials, as will most common QMCPACK workloads, and
they allow the effects of the changing electron count to be addressed
in a direct manner.  However, we will refer to the entire set where
appropriate to demonstrate the universality of the algorithm across
problem types.

\section{System details}\label{sec:sys}

We used two different shared memory multi/many-core processors to capture
performance evolution at each major step:  i) dual socket \xeonReg E5v4 CPU
(BDW) and ii) second generation \xphiTMKNL 7250P (KNL).  We also use IBM
Blue Gene/Q (BG/Q) processor for demonstrating portability of our
performance improvements.  Two types of systems are used for multi-node
scaling and performance analysis: i) Trinity at Los Alamos National
Laboratory with KNL processors and Cray Aries Dragonfly interconnect; ii) Serrano
cluster at Sandia National Laboratories with dual-socket BDW processors and
Intel\registered Omni-Path interconnect.

Two different BDW SKUs are used: i) 20-core single socket E5-2698 v4 CPU for
the single-node performance analysis, and ii) 18-core dual socket E5-2695 v4
for multi-node runs on Serrano cluster.  KNL processor is used in Quad cluster
mode and wherever possible, KNL-MCDRAM is used in flat mode.  For a few runs,
memory footprints exceed 16GB MCDRAM capacity; those are done in MCDRAM-cache
mode on KNL.  We use 64 out of 68 cores on the KNL machine, leaving a few cores
out to do OS related tasks.  Performance comparisons are done between a KNL node and
single-socket of a dual-socket BDW node considering their power budgets and NUMA
characteristics. 

We use tools\footnote{\CompilerDisclaimer} from \studioReg~\cite{parallelStudioCite} on Intel\registered
platforms.  BDW and KNL use architecture specific compiler
options~\cite{intelComp}.  For advanced hot-spot profiling, \vtuneTM
(VTune)\cite{vtuneCite} was utilized.  Roofline performance analysis was done
with an engineering version based on Intel\registered Advisor 2017 update
2~\cite{advisorCite}.  On BG/Q, we used Clang compiler version 4.0.0 (bgclang
r284961-stable)~\cite{clang}.

\section{Reference QMCPACK}

The baseline of this work uses the latest public release of QMCPACK
3.0.0~\cite{qmcpackSite} with the mixed precision feature turned off.  
This section describes the reference QMCPACK
implementation and presents an analysis of its performance.

\subsection{Baseline with AoS data types}

Figure~\ref{code:base} presents a simplified QMC code, containing a driver
method \code{psuedo\_qmc} and core abstractions for D-dimensional particle
simulations.  It is constructed to mimic the structure of QMCPACK 3.0.0.  The
threading is implemented with OpenMP.  \code{ParticleSet} and
\code{TrialWaveFunction}, the main compute objects, are created per thread as
denoted by \code{E\_th} and \code{Psi\_th}. Here \code{nw} is a dynamic variable
during a DMC run which represents number of walkers.  
It is updated during the reweight and branch walkers step
(L13 in Alg.~\ref{alg:DMC}).  A generic \code{Vector<G>} is used to represent
any attribute such as positions.  The most basic and important attribute
\code{R} encapsulates the positions of $N$ particles in an AoS type,
\code{Vector<TinyVector<T,D>>}.

\begin{figure}
\centering
\begin{minipage}{.42\textwidth}
\lstset{language=c++,
  basicstyle=\ttfamily\scriptsize,
  commentstyle=\itshape\color{blue},
  breaklines=true,
  numbers=left,
  frame=single,
  captionpos=b
}
\begin{lstlisting}
//fixed D-dimensional vector for each particle
template<typename T, unsigned D>
class TinyVector { T X[D]; };

//generic 1D container
template<typename G> class Vector{
  std::vector<G> X; 
};

//Walker class
template<typename T, unsigned D>
class Walker{
  Vector<TinyVector<T,D>> R;//positions (AoS)
  Buffer<T> Any; //anonymous buffer
};

template<typename T, unsigned D>
class ParticleSet{
  using Walker_t=Walker<T,D>;
  //Arrays of particle attributes
  Vector<TinyVector<T,D>> R;//positions (AoS)
  Vector<TinyVector<T,D>> G;//gradients (AoS)
  Vector<T>               L;//laplacians

  //containers of Walkers
  Vector<Walker_t*>  Walkers;

  //copy a Walker to perform a MC step
  void loadWalker(const Walker_t& awalker) {
    R=awalker.R;
  }
};

void pseudo_qmc() {
   using Particles=ParticleSet<double,3>;
   Particles E;
   Particles Ions; //shared among threads
   TrialWaveFunction Psi(E,Ions); 
   #pragma omp parallel
   {
     Particles E_th(E);
     TrialWaveFunction Psi_th(Psi)
     #pragma omp for nowait
     for(size_t iw=0; iw<nw; ++iw) {
       E_th.loadWalker(*(E.Walkers[iw]));
       for(size_t k=0; k<N; ++k) {
         //PbyP update with DMC Algo.1
       }
       E_th.storeWalker(*(E.Walkers[iw]));
     }
   }
}
\end{lstlisting}
\end{minipage}
\caption{
  A simplified QMC code using OpenMP, showing a driver method \code{psuedo\_qmc} and
  core abstractions for D-dimensional particle simulations. Operators 
  and other utility methods are not shown. \label{code:base}}
\end{figure}

A \code{Walker} object is a simple container to manage the positions, physical
quantities such as $E_L$, weight, age etc and an anonymous \code{Buffer} to
store internal state for fast PbyP updates.  The exact form and the composition
of $\Psi_T$ is only known at run time and each orbital component can have any
number of scalars to compute the differences before and after a move.
\code{loadWalker/storeWalker} methods copy a \code{Walker} data to the compute
objects for independent updates on a block of \code{Walker}s.  High-level
physics is expressed using only \code{ParticleSet} and
\code{TrialWaveFunction}.

The reference implementation pre-computes and stores all the elements needed by
\code{TrialWaveFunction} for the PbyP updates beforehand and then retrieves and
modifies them during the updates. The anonymous \code{Buffer} holds any number
of scalars to reconstruct the complete state of a \code{Walker}
without recomputing.  The memory-demanding J2 (eq. \ref{eq:Jratios}) keeps full $N$-by-$N$ matrices
for $U_2(i,j), \nabla U_2(i,j)$ (3D vector) and $\nabla^2 U_2(i,j)$ and uses
minimum $5 N^2$\code{sizeof(T)} per \code{Walker}.  This store over compute policy was
adopted when the FLOPS (sqrt, inverse, sincos) were expensive compared to
reading/writing to a memory region and the number of cores per node was small
(16 on BG/Q). 

\subsection{Performance analysis of baseline}

A DMC run performs many steps $M\sim 10^6$. Either the total execution time
$T_{\rm CPU}$ or the throughput which is equal to the number of MC samples generated per second, can
be used as the figure of merit.  For these benchmarks, we use 100-1000 steps to
make runs manageable and compute the throughput as $P=M \left<N_w\right>/T_{\rm CPU}$. 
Here, $\left<N_w\right>$ denotes the average $N_w$.
This
throughput is
representative of the production runs of the same target population and is
directly correlated to the DMC efficiency $\kappa$.  Ratios of throughputs are
used to show the relative performance of the different runs on a given system
and the runs on multiple systems.  

For the baseline (Ref), all the quantities are in double precision by compiling
QMCPACK with \code{QMC\_MIXED\_PRECISION=0}, except for the Bspline-SPO
(Bspline-v and Bspline-vgh) in Fig.~\ref{fig:prof}. This was the standard for
production calculations prior to the version 3 release.  We show performance
improvements in two steps: (Ref+MP) uses mixed-precision with the reference
code and (Current), mixed-precision  with the final optimized code which includes all the
techniques described in Sec.~\ref{sec:methods}. Other intermediate steps are
not presented but can be measured using different build options and miniapps.  

The Ref profiles for the NiO benchmarks on KNL in
Fig.~\ref{fig:prof} reveal that the computations of the distance relations
among electrons (AA type) and between electrons and ions (AB type) and J2 make
up close to 50\% of a run.  This is in contrast to the earlier profile of a
smaller problem on older Harpertown quad-core processor that shows close to
50\% is spent on Bspline-SPO routines~\cite{qmcpackGPU}.  We attribute these
changes in the profiles to i) an increasing penalty of scalar operations using AoS
data types on wide SIMD processors, ii) the high pressure on memory subsystems
with more electrons in our larger benchmarks and iii) optimization of
Bspline-SPO evaluations by converting critical calculations to single
precision.  Future problems are more demanding and the current implementation
does not provide sufficient performance for practical QMC simulations of
large-scale problems we would like to tackle in the future such as a disordered
1024 atom supercell of NiO.

\section{Transforming QMCPACK}\label{sec:methods}

In order to address the performance bottlenecks identified previously we take a
multi-step approach.  First, we create miniapps upon which to test our
algorithmic improvements.  Next we change the data layout in many sections of
the code, increase the use of single precision computations, and finally
overhaul many distance table based algorithms involved in various parts of the
code.  In this section we describe these optimizations, focusing on
both the methodology as well as the reasons behind the chosen algorithms.

\subsection{Miniapps}

We created a set of miniapps to explore solutions for the three main classes
responsible for the hot-spots separately: DistTable, Jastrow (J1 and J2) and Bspline-SPO.
Finally, \code{miniQMC} tests all the three main components.  Each miniapp
mimics a QMC calculation using PbyP update and non-local pseudopotentials as
shown in Alg.~\ref{alg:DMC} and Fig.~\ref{code:base}.  They reproduce the computational patterns, memory
use, data access and thread-level parallelism of the production QMC code as
realistically as possible.  Command-line options are used to change the
problems ($N$, the cutoff radius and etc) for fast prototyping, debugging and
analysis.  

These miniapps allow us to explore a large design space without global code
modifications and to quantify any impact of the new implementations before
complete integration.  We use the performance model based on the theoretical
analysis of QMC algorithms and empirical data on multiple platforms to project
the productivity gains in real QMC simulation environments. Once we narrow down
the solution space in miniapps, we implement the new data-types and methods in
QMCPACK to maximize the reuse of the existing framework and to continue
supporting the high-level physics abstractions that are essential for QMC
method development.

\subsection{Mixed precision}

For the first optimization of the code, our work expands the use of single
precision in the most performance critical kernels of QMCPACK including
DistTable and Jastrow.  We convert the key data structures and
calculations to single precision, while keeping the precision-critical
computation in double precision.  These improvements are already available in
3.0.0 version and are enabled with \code{QMC\_MIXED\_PRECISION=1} flag.  Prior
to v3.0.0, only Bspline-SPO data and evaluations use single precision.  This
new feature significantly speeds up computations and reduces the memory usage
associated with walkers and threads by half. 

\subsection{SoA data layout update}

The object-oriented (OO) and generic programming paradigm is widely adopted for
large-scale, complex HPC applications such as QMCPACK.  The AoS datatypes (C++
objects) are natural choices to express mathematical concepts and logics of
physics simulations.  Expression templates allow optimization of complex
algorithms at the compiler time.  However, the abstraction penalty can be high
and outweighs the benefit of using OO and generic approaches.  

Production applications, such as QMCPACK, are complex often comprising millions
of lines of code and support various needs of the developers and users. The
SIMD-friendly solutions to accelerate the current hot-spots may harm the
overall performance and limit how the high-level physics is expressed. It is
essential to consider the algorithms and the balance of computations and memory
access of the entire application.  Any changes must be portable and extensible
by the developers.  This work adheres to C++11 and OpenMP 4 SIMD standards to
facilitate auto vectorization and optimizations by compilers and does not use
any platform-specific optimizations, although they are not excluded for the
future development.

\begin{figure}
\centering
\begin{minipage}{.42\textwidth}
\lstset{language=c++,
  basicstyle=\ttfamily\scriptsize,
  commentstyle=\itshape\color{blue},
  breaklines=true,
  numbers=left,
  frame=single,
  captionpos=b
}
\begin{lstlisting}
//Generic SoA container and key operators
template<typename T, unsigned D>
class VectorSoaContainer{
  aligned_vector<T> X; 
  TinyVector<T,D> operator[](size_t i) const;
  template<typename VA> 
  VectorSoaContainer<T,D>& operator=(const VA& rhs);
};

//Add a new data member Rsoa in SoA
template<typename T, unsigned D>
class ParticleSet{

  VectorSoaContainer<T,D> Rsoa;

  void loadWalker(const Walker_t& awalker) {
    R=awalker.R; 
    Rsoa=awalker.R; //AoS-to-SoA assignment
  }
};
\end{lstlisting}
\end{minipage}
\caption{
  Update to simplified QMC code in Fig.~\ref{code:base} with
\code{VectorSoaContainer} and modified \code{ParticleSet} class. Utility functions are not shown.\label{code:vsc}}
\end{figure}

In order to enhance SIMD efficiency, we adopt the concepts and techniques of
SDLT~\cite{sdlt,sdltqlib} and implement a generic container, \code{VectorSoaContainer<T,D>} (VSC), to
encapsulate a vector of non-scalar types.  A VSC object is a transposed form of
the corresponding AoS object in SoA format and provides access and utility
methods to interact with the AoS counterparts in place.  The SoA containers use
cache-aligned allocators chosen at the compiler time.  On Intel platforms, we
use the TBB cache-aligned allocator as their default allocator.
Figure~\ref{code:vsc} presents important details of \code{VectorSoAContainer}
class and its typical use in QMCPACK.

We introduce a new data member \code{Rsoa} in the \code{ParticleSet} class and
implement SIMD-friendly methods in DistTable and Jastrow classes using the new
SoA objects. The overhead of the duplicate containers to hold electron
positions in \code{R[N][3]} (AoS) and \code{Rsoa[3][Np]} (SoA) is negligible in
terms of computation and storage. Here, \code{Np} includes the padding for
alignment. The only extra operations are the additional assignment in
\code{loadWalker} and the update when a move is accepted.

For the $k$-th electron move during the PbyP update (L47 Fig.~\ref{code:base}),
all the routines are functions of the position \code{R[k]} of the active
electron $k$ and \code{Rsoa} of the electrons and ions.  The computational kernels
are expressed as 1-by-$N$ and 1-by-$N_{\rm ion}$ relations, \eg the distances
$d(k,i)=|\vecr_i-\vecr_k|$, displacement vectors $d\vecr(k,i)=\vecr_i-\vecr_k$
and $\nabla U_2(k,i)$ for $i\ne k$. They are now implemented using the loops
over $N$ or $N_{\rm ion}$ that can be easily vectorized by the compilers.  When a
move is accepted, both \code{R} and \code{Rsoa} (6 floats), are updated with the new
position for the active electron.  The positions of the ions
(a \code{ParticleSet} object) are fixed during a QMC calculation and the
ions' \code{Rsoa} is reused throughout the calculation.

\subsection{Forward update method}

Use of miniapps for the development has additional advantages over tackling the
full code transformation from either the top or bottom. They expose the
performance bottlenecks and inefficiencies of the current implementation that
were not obvious or hidden by the primary hot-spots analysis.  As we improve the SIMD
efficiency with SoA containers, the cost of memory movement and the
pre-compute-and-store policies turned out to be too high and diminish the benefit
of the AoS-to-SoA layout transformations in the main computations.  

\begin{figure}
  \centering
  \includegraphics[scale=0.4]{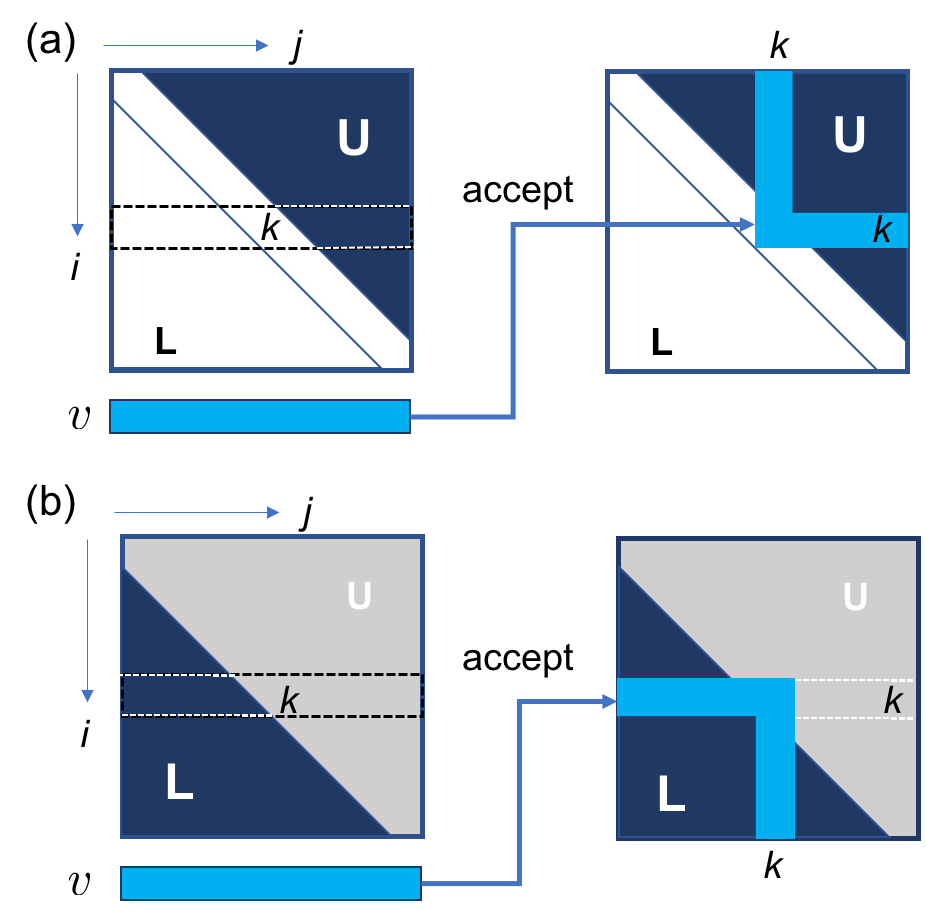}
  \caption{Schematics for AA (symmetric) distance table management (a) before and (b) after the optimizations.
  $v$ is a separate array to hold the temporary data for the $k$-th electron move.
  The column update in (b) is later removed with the compute on the fly optimization.
}
  \label{fig:fastalgo}
\end{figure}

As a particle-based method, managing the distance tables, equivalent to the
nearest-neighbor lists in classical molecular dynamics codes, is critical for
efficiency.  The distance-table objects can be reused any number of times
depending upon how many Jastrow orbitals constitute the trial wavefunction
$\Psi_T$ and what basis set is used for SPOs. They are also used by Hamiltonian
objects when the measurements are made.

The top panel of Fig.~\ref{fig:fastalgo} illustrates how the reference QMCPACK
handles the electron-electron (AA) distance table.  It stores the upper
triangle in a packed storage as shown in dark blue and labeled as U.  When the
$k$-move is accepted, the temporary container $v$ is copied to update U.  The
packed storage needs $N(N-1)/2$ scalars and requires only $N$ copies for the
update.  However, the access patterns on SIMD processors are not favorable for
compiler auto-vectorization due to unaligned accesses.  Such inefficient
data-access patterns are repeated in other routines.

We develop new algorithms to reduce memory operations, exploiting the
sequential nature of the PbyP update: it performs ordered moves of $N$
electrons and most of the properties associated with $k^{\prime}<k$ electrons
are not needed during the drift-and-diffusion stage.  Instead of updating the
full row or column of the active electron $k$, we update only the data
necessary for the future moves and delay other computations and retain minimum
quantities in memory, until the measurement is made. 

The bottom of
Fig.~\ref{fig:fastalgo} presents the new method for the AA distance table.  We
use the full $N\times N^p$ storage (including padding) even for the AA types,
increasing the memory use roughly by two.  This compromise is justified as we
can achieve close to the ideal speedup of vectorization from double scalars to
packed floats, with cache-aligned access for each row of $N^p$ and $v$.
We develop a forward-update method, leaving U untouched or partially
updated as the Upper triangle is not used by other methods.  The $k$-th column
update is strided by $N^p$ but only $k^{\prime}>k$ are updated upon acceptance,
leaving the number of copy operations unchanged.

Similar approaches are taken in electron-ion (AB) distance table and Jastrow
orbitals to eliminate unnecessary memory movements.  The bulk of improvements
in the Jastrow routines comes automatically with the changes in the distance
tables and \code{ParticleSet}. Jastrow orbitals are the consumers of AA or AB
distance tables and the cache-aligned, SIMD-friendly data-types allow
straightforward code modifications and facilitate compilers' autovectorization. 

\subsection{Compute on the fly and memory savings}

In addition, new algorithms~\cite{JastrowAlgo} are developed to reduce memory
used by Jastrow objects.  The factorized form of $\Psi_T$ makes the ratio
computations a product of each component. The J1 contribution to the ratio
Eq.~(\ref{eq:Jratios}) depends on the difference of $U_k^1$ for the $k$-th
electron before and after the move, as
\begin{equation}
  \Delta J_1 = U^{1}_k(\textbf{R}^{\prime})-U^{1}_k(\textbf{R}),\; U_k^1=\sum_I^{N_{\rm ion}} U_I(|\vecr_I-\vecr_k|).
\end{equation}
Two-body Jastrow takes the similar form as $\Delta J_2=U^{2}_k(\textbf{R}^{\prime})-U^{2}_k(\textbf{R})$.  
To reuse the
computed values, the Ref implementation uses three $N\times N$ matrices for the
values, gradients (D=3) and Laplacians, total of $5N^2$ scalars per walker. The
gradients are stored in an AoS container and both the column and row are
updated for each accepted move. The SoA transformation in DistTable
objects makes the Jastrow loops highly vectorizable by the compilers.
With highly sped up
computations, due to the single-precision use and SoA transformations, we can
afford to eliminate the intermediate data all together and keep the memory use
of J2 at $5 N$\code{sizeof(T)}. We apply such compute-on-the-fly approaches whenever
profitable.

Finally, we redesign \code{ParticleSet} and \code{TrialWaveFunction} member
functions to clearly define the roles and requirements of the virtual functions
for move, accept/reject and measurement. These changes make it possible to
expand compute-on-the-fly methods to DistTable.  Instead of precomputing the full
table and updating the column and row as depicted in Fig.~\ref{fig:fastalgo},
we now compute the row $k$ with the current position $\vecr_k$ before making
the move. 
This eliminates the strided copy for the column updates.  We retain
$O(N^2)$ storage in DistTable, since they are used multiple times by
\code{Hamiltonian} objects.  The results denoted as ``Current'' are obtained
using the implementation that includes all the optimization steps discussed in
this section.

\section{Results and discussions}

Figure~\ref{fig:scaling} summarizes the outcome of the transformative changes
of this work\footnote{\BenchmarkDisclaimer}.  We present the relative performance of four sets of
strong-scaling runs of a 64-atom supercell of NiO on Trinity (KNL) at LANL and Serrano
(BDW) systems at SNL. The throughputs are normalized by that of
the Ref code on BDW using 64 sockets (32 nodes and 1152 cores). We use 1 MPI
task per KNL node (BDW socket) and two threads per core.  The target DMC
population is set at 131072.  This corresponds to one walker per thread, on
average, for the 1024-node runs on KNL. 
In all cases, the parallel efficiency is high, 90\%~(KNL) and 98\%~(BDW),
and 2-4.5x speedup is obtained through the optimizations.

All the performance improvements in Fig.~\ref{fig:scaling} are attributed to
the data-layout transformations, reduced memory operations and the expanded use
of single precision.  The MPI communications are the same for both Ref and
Current code: allreduce to compute running averages for $E_L$ and other global
properties and send/recv of serialized \code{Walker} objects during the
load-balancing steps.  The memory-reduction algorithms in Jastrow reduce the
\code{Walker} message size by 22.5 MB for the NiO-64 problem.  There is,
however, no fundamental change in communications that still have low overhead.
Therefore, we focus on the single-node benchmarks and provide comprehensive
analysis of the on-node performance to show the impact of our work on
performance, memory footprint and energy consumption for the rest of the
section.

\subsection{Roofline and hot-spot analysis}

\begin{figure}
  \includegraphics[scale=0.38]{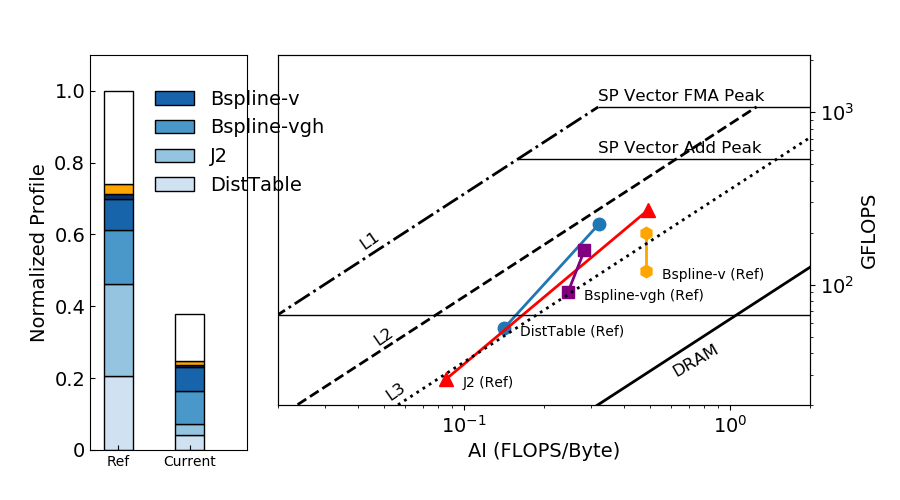}
  \caption{Normalized hot-spot profile and roofline analysis of NiO-32 on BDW.
   Current version hot-spot profile accomodates the speedup wrt.
Ref version. 
  }
  \label{fig:bdwrl}
\end{figure}

The hot-spot profile and roofline~\cite{roofline,cacheRoofline} analysis of the
32-atom NiO supercell in Fig.~\ref{fig:bdwrl} shows the evolution of the four
major kernels before and after the optimizations on BDW.  Similarly,
Fig.~\ref{fig:prof} shows hot-spot profiles of NiO benchmarks on KNL. The
transformations significantly decrease the time spent in DistTable, J2 and
Bspline-vgh.  Other determinant-related computations, SPO-vgl and DetUpdate,
are sped up by more than two with the double-to-single transition in $A^{-1}$.
The roofline performance model on BDW shows large jump in both AI and FLOPS
with the Current code.  Efficient vectorization is enabled in DistTable,
Jastrow, Bspline-vgh and SPO-vgl with the SoA data-types.  The greater increase
in AI and FLOPS of DistTable and J2 is the combined effect of the expanded use
of single precision and the improved data structures and algorithms.  Bspline-v
kernel is unchanged but its efficiency increases with the memory optimizations
on BDW.

Our optimizations have quantitatively different effects on BDW and KNL
processors due to their architectural differences.  KNL has twice the single
precision SIMD width of BDW's, making the theoretical vectorization speedup
twice as large.  The bandwidth of 16 GB MCDRAM in flat mode is about 8 times
higher than that of one-socket BDW. The cache subsystems, their sizes and
associativities, are also different.  The shared L3 cache on BDW can make up
for the low DDR bandwidth: Fig.~\ref{fig:bdwrl} shows that all four kernels lie
above the L3 roofline after the optimizations. Despite these architectural differences,
qualitative impacts of the optimizations are the same for both processors.


The data-layout transformation enables close to the ideal speedup in DistTable
computations, due to its contiguous stream of data access. For Jastrow
routines, the vectorization efficiency is slightly lower due to the branch
conditions originated from the finite cutoff of the Jastrow functors in
Fig.~\ref{fig:nioj}.  Compute-on-the-fly policy in Jastrow routines are
critical as we eliminate all $\mathcal{O}(N^2)$ memory storage.  The only
remaining $\mathcal{O}(N^2)$ storage per walker comes from the determinant
objects in storing $A^{-1}$.  All optimizations of Current work result in
5x~(DistTable), 8x~(Jastrow), 1.7x~(Bspline-vgh) and 1.3x~(Bspline-v) speedups
for the NiO-32 benchmark on BDW.  Figure~\ref{fig:prof} shows
the normalized profiles of both NiO benchmarks on KNL, showing similar speedups
for each routine.

\subsection{Benchmark results and discussion}

Having established the efficiency and scalability of QMCPACK with our current
methods, we turn to the detailed performance analysis of NiO benchmarks on
single KNL and BDW processors.  The system details are provided in Sec.~\ref{sec:sys}.  To keep
the the amount of work similar, we use the target population of 1024 (KNL) and
1040 (BDW), equivalent to 8 and 24 \textit{average} walkers per thread on KNL
and BDW, respectively.

Our hyperthreading study of the 32-atom NiO supercell benchmark with the
optimized Current version shows its positive impact on both BDW and KNL
processors.  This is expected because hyperthreading can hide latency in
memory-intensive operations such as Bspline-SPO routines. They are memory-latency
sensitive due to random accesses of the 4-dimensional read-only table and are
also memory-bandwidth limited.  Using 2 threads per core provides 10\% and
8.5\% throughput improvements for BDW and KNL respectively.  On KNL, using 2
threads per core is optimal for this system and using 3 or 4 threads per core
does not improve the throughput.  This is generally true for other problems we
have investigated.

\begin{figure}
  \centering
  \includegraphics[scale=0.5]{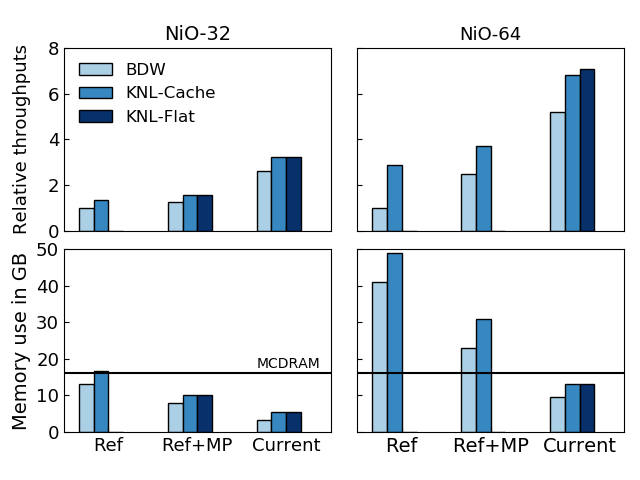}
  \caption{Speedup and memory-usage reduction of NiO benchmarks.  The
  throughputs are normalized by those of Ref on BDW.}
  \label{fig:nioKNLSpeedup}
\end{figure}

Figure~\ref{fig:nioKNLSpeedup} shows performance and memory usage of NiO on BDW
and KNL in cache and flat memory modes for the Ref, Ref+MP and Current.  The
missing data of KNL-flat is due to the memory footprint being more than 16 GB,
the capacity of the MCDRAM.  The throughputs are normalized by the Ref on BDW
and higher throughput means higher performance. 

Mixed-precision implementation (Ref+MP) reduces memory bandwidth usage by
storing the key datasets in single precision and accelerates bandwidth-bound
routines.  However, it does not benefit functions with low SIMD
efficiency, limiting its impact on KNL.  The 64-atom supercell of NiO doubles
the problem in size and therefore, its computational cost and memory use are accordingly
higher. It is expected to be bandwidth bound and gains more by MP than smaller
problems. The speedups on KNL, 1.3x of NiO-64 compared 1.16x of
NiO-32 support this performance projection.  
Expanded use of single precision, together with the help of shared L3 on BDW, 
lowers the bandwidth demands for both the
benchmarks and leads to higher speedups, 
2.5x~(NiO-64) and 1.3x~(NiO-32).

The performance of Current runs are more than doubled on both BDW and KNL
compared to Ref+MP. More importantly, the memory usage has gone down
dramatically as much as 36 GB from Ref for the NiO-64 benchmark, allowing
all the benchmarks to run on KNL in flat mode. The performance gains from cache
to flat mode are modest, around 3\% for NiO-64. The importance of high BW is
evident for the NiO benchmarks.  Exclusive use of DDR (\code{numactl -m 0})
slows down the Current by 5.4x for NiO-64, which is commensurate with the
stream bandwidth difference of MCDRAM and DDR on KNL. The low BW of DDR affects
the 32-atom supercell of NiO less, slowing it down only by 2.3x, as the
compute-bound routines play greater roles for the smaller problems.

The bottom of Fig.~\ref{fig:nioKNLSpeedup} shows the measured memory usage in
GB on BDW and KNL for NiO problems.  The memory footprint of Ref QMCPACK grows
as $\gamma (N_{th}+N_w)N^2$ excluding the read-only 3D B-spline table which is
shared by all the threads. 
Here $N_{th}$ and $N_w$ represent the number of threads and walkers respectively.
The pre-factor $\gamma$ depends on the details of
$\Psi_T$ and the minimum is 60 bytes to store J2 and determinant objects in
double precision.  This allocation policy is the design choice to make
thread-level parallelization efficient by maximizing the data locality and
removing data racing conditions.  Separating \code{Walker}s from the compute
engines, \code{ParticleSet} and \code{TrialWaveFunction}, makes it possible to
use an arbitrary number of \code{Walker}s per node and any number of nodes
under the memory constraints. 

\begin{figure}
  \centering
  \includegraphics[scale=0.4]{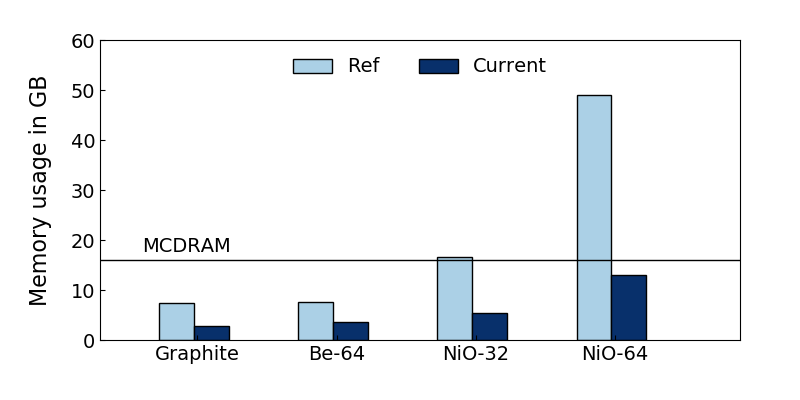}
  \caption{Memory usage on KNL processor.}
  \label{fig:allKNLMemory}
\end{figure}

The increased thread-level parallelisms on newer processors, such as KNL, limit
the problems we can solve using the Ref implementation.  We expect that the
simulations of 1000s of electrons will become the norm, rather than an unusual
scenario in the near future. Reducing memory footprint is critical, while
utilizing all the resources available on a node for the productivity.
Figure~\ref{fig:allKNLMemory} shows
$\mathcal{O}(N^2)$ memory savings on the four benchmarks in Current
through the use of new algorithms and
expanded use of single precision. For instance, 36 GB reduction in memory is
achieved for NiO-64 and the total memory footprint is less than 16 GB, the
memory capacity of a BG/Q node.  Such savings open up new opportunities for the
scientists and allow them to study the problems they cannot solve with Ref
QMCPACK today.

\begin{figure}
  \centering
  \includegraphics[scale=0.32]{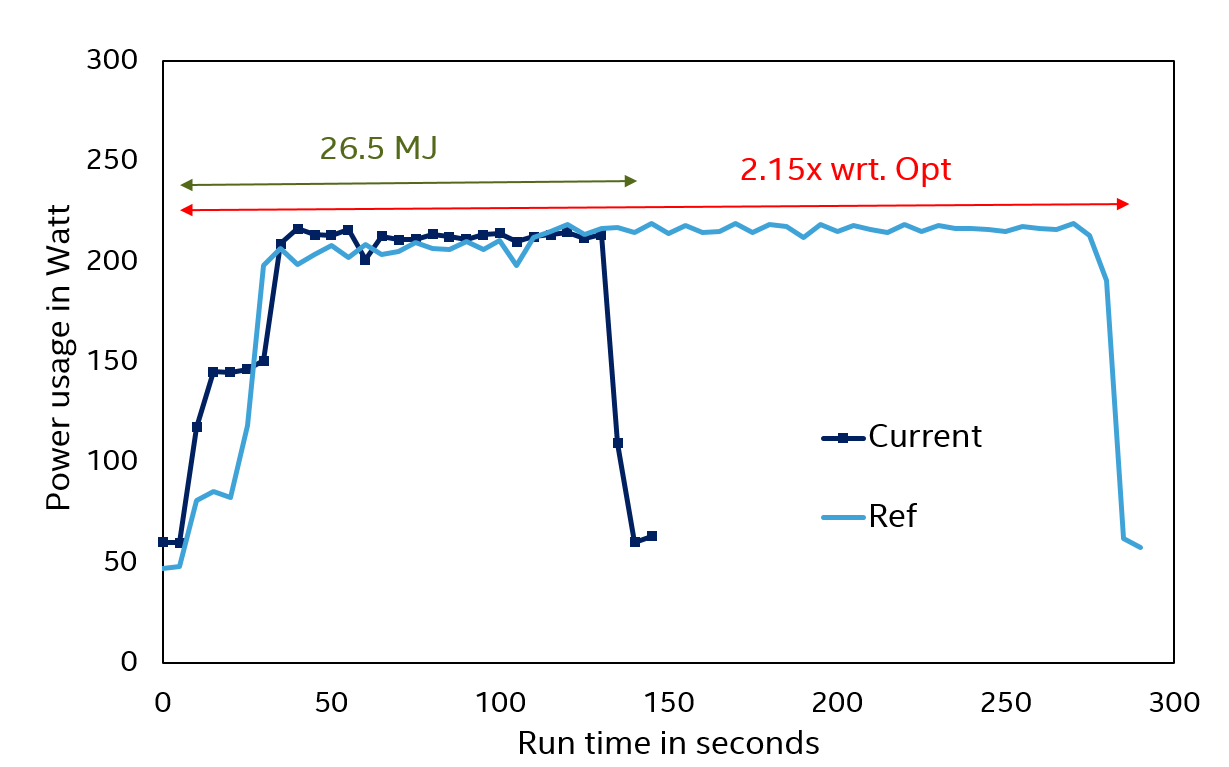}
  \caption{Energy usage of NiO-32 benchmark on KNL.}
  \label{fig:powerComp}
\end{figure}

Figure~\ref{fig:powerComp} shows energy reduction after the optimizations
(Current) compared to Ref for NiO-32.  Power usage is plotted
against the time of execution.  The Ref version required to use MCDRAM in
cache mode as mentioned earlier.  Power is measured with the turbostat Linux
utility with a 5 second interval. We add PkgWatt and RAMWatt values, which
represent the total power used by the CPU+MCDRAM and DDR.  Power usage
fluctuate within the range of 210-215 watt during the DMC phase for both Ref
and Current.  A similar power profile was obtained for the larger NiO-64
problem on KNL.  Excluding the initialization and warmup time, the energy 
reduction is roughly equal to the speedup obtained with the optimizations.
This means huge energy savings for the production simulations running on 1000's
of nodes for hours and days as well as huge productivity gains in science.

\begin{table}[!b]
\caption{Speedup of Current over Ref on BG/Q, BDW and KNL, respectively.}
\label{tab:allSpeedup}
\centering
\begin{tabular}{|l|c|c|c|c|}
\hline
& Graphite & Be-64 & NiO-32 & NiO-64\\
\hline
BG/Q & 1.6 & 1.3  & 1.3  &  2.4\\
\hline
BDW & 2.9  &  3.4  &  2.6  &   5.2\\
\hline
KNL & 2.2  &  2.9  &  2.4  &   2.4\\
\hline
\end{tabular}
\end{table}

\subsection{Performance summary and portability}

The improved performance from our work is not just limited to BDW and KNL.  As
pointed out, no platform-specific optimizations or intrinsics are used in
Current and we employ the standard features modern C++ compilers support.
Table~\ref{tab:allSpeedup} gives the final speedups of the four benchmarks on
BG/Q, BDW, and KNL processors.  The speedup compares the performance of the
Current implementation over the Ref code on each system and does not reflect
the absolute performance of different processors.  These benchmarks are
distinct in their sizes and computational characteristics due to the different
constituent ions and the cell shapes. They exercise different code paths
determined at run time for each benchmark. The compiler's support for C++11,
OpenMP SIMD and their abilities to produce optimized binaries vary. Nevertheless,
we are able to accelerate the entire QMC simulations across the platforms and
the problems.

\subsection{Outlook and future work}

The much improved efficiencies of the top hot-spots increase the importance of
the other kernels that have not been addressed so far, including
DetUpdate for $A^{-1}$ update.  Figure~\ref{fig:prof} shows
DetUpdate is 10~\% for NiO-64 using Current, as opposed to 7~\% with Ref.
The asymptotic $\mathcal{O}(N^3)$-scaling of QMC methods arises from
DetUpdate based on Sherman-Morrison formula.  For the current problems
on CPUs with multiple cache levels and ample capacity, the computations are
dominated by DistTable, Jastrow, and SPO evaluations and grow as
$\mathcal{O}(N^2)$.  However, as the system size grows, DetUpdate using
BLAS2 becomes increasingly important and becomes
the bottleneck of QMC calculations.

Several alternatives based on Woodbury matrix identity~\cite{woodbury}, the
generalization of Sherman-Morrison formula, can be applied to DetUpdate.  One
promising solution is a \textit{delayed-update} scheme designed to evaluate
multiple accepted moves before any updates are made to $A^{-1}$~\cite{detupdategpu}.
The delay factor can be adjusted to optimize the performance of higher BLAS
functions for the update and the ratio computations for any size $N$. No
structural changes are required to implement these new DetUpdate methods.

The efficient vectorization and reduction in memory footprint, \eg to run
NiO-64 using 128 threads entirely on 16 GB MCDRAM in flat KNL memory mode, is
critical to solve today's problems faster and to enable simulations of much
bigger and demanding future problems.  The transformations presented in this
work increase the science productivity and resource utilization of the systems
we have and are the critical step to future-proof QMCPACK for the systems we
will have.

Let's consider a 512-atom supercell of NiO (6144 electrons).  It is 8 times
bigger than the current NiO-64 and would take 512 times longer per step and require 64
times more memory with the Current, even with all the optimizations.  Exposing
extra parallelisms is the only path to make QMC practical to study such
problems.  BLAS3 routines are highly optimized and parallelized on any platform.  The
``fat'' loops over the electrons and ions are ideally suited to parallelize the
computations for each walker.  

Our previous work~\cite{qmcipdps17} demonstrated that tiling of the big
B-spline table and parallel execution over the array-of-SoA (AoSoA) objects can reduce
the time to complete a QMC step. We propose to extend those ideas to full
QMCPACK.  Its object-oriented designs are amenable for either nested loop
parallelization or any task-based parallelism. OpenMP standards support
various ways to implement the parallel executions.  Which solution will provide
the most productive path for the science is unknown. We expect our approaches
based on miniapps and iterative transformation processes to facilitate future
developments as well.

\section{Conclusions}

We presented single-node optimizations for QMCPACK, a leading US-DOE quantum Monte
Carlo application, and demonstrated the transferability of these optimizations
to highly parallel runs on multiple platforms.  A set of miniapps representing
the computational and data access patterns of QMC were developed for fast
exploration, debugging and evaluations. We applied the structural changes in
QMCPACK by introducing new abstractions in the SoA format and implementing the
methods that can be optimized by modern C++ compilers.  Our work systematically
expanded the use of single precision to reduce memory bandwidth demands and
footprint, while preserving the fidelity of double-precision calculations.
Taking advantage of the increased SIMD efficiency of the new kernels, we
developed and implemented new algorithms to further improve the performance
and to reduce the memory footprint. All these are
seamlessly incorporated in the production QMCPACK in portable and maintainable
ways to increase the productivity of the developers and users.

\section*{Acknowledgment}
We thank Jason Sewall, Roland Schulz, Victor Lee, John Pennycook,
and Dayle Smith for their helpful
discussions and reviewing this manuscript.
We also thank Intel\registered Advisor team for providing timely engineering builds and quick responses.
This work is supported by Intel Corporation to establish the
Intel Parallel Computing Center at Argonne National Laboratory.  LS was
supported by the Advanced Simulation and Computing - Physics and Engineering
models program at Sandia National Laboratories.  AB was supported through the
Predictive Theory and Modeling for Materials and Chemical Science program by
the Office of Basic Energy Science (BES), Department of Energy (DOE).  Sandia
National Laboratories is a multi-program laboratory managed and operated by
Sandia Corporation, a wholly owned subsidiary of Lockheed Martin Corporation,
for the U.S. Department of Energy's National Nuclear Security Administration
under Contract No. DE-AC04-94AL85000. This research used resources of the
Argonne Leadership Computing Facility, which is a DOE Office of Science User
Facility supported under Contract No. DE-AC02-06CH11357.
This research was supported by the Exascale Computing Project (17-SC-20-SC), 
a collaborative effort of the U.S. Department of Energy Office of Science and the National Nuclear Security Administration.

\bibliographystyle{ACM-Reference-Format}
\bibliography{qmcpack}

\end{document}